\documentclass{article}

\usepackage{microtype}
\usepackage{graphicx}
\usepackage{subfigure}
\usepackage{booktabs} 
\usepackage[most]{tcolorbox}

\usepackage{colortbl}

\definecolor{boxcolor1}{RGB}{230,240,255} 

\newtcolorbox{questionbox}[1][]{
  colback=#1,
  colframe=black,
  boxrule=1pt,
  arc=5pt,
  outer arc=5pt,
  enhanced,
  breakable
}

\usepackage{hyperref}

\usepackage[accepted]{ml4astro2025}

\usepackage{amsmath}
\usepackage{amssymb}
\usepackage{mathtools}
\usepackage{amsthm}

\usepackage[capitalize,noabbrev]{cleveref}

\theoremstyle{plain}

\theoremstyle{definition}

\theoremstyle{remark}

\usepackage[textsize=tiny]{todonotes}

\begin{document}

\twocolumn[
\mlforastrotitle{Teaching LLMs to Speak Spectroscopy}
\mlforastrotitlerunning{Teaching LLMs to Speak Spectroscopy}




\begin{mlforastroauthorlist}
\mlforastroauthor{Nesar Ramachandra}{cps}
\mlforastroauthor{Yuan-Sen Ting}{comp,sch} 
\mlforastroauthor{Zechang Sun}{tsinghua}
\mlforastroauthor{Azton Wells}{cps}
\mlforastroauthor{Salman Habib}{cps}
\end{mlforastroauthorlist}

\mlforastroaffiliation{cps}{Computational Science Division, Argonne National Laboratory, 9700 South Cass Avenue, Lemont, IL, USA}
\mlforastroaffiliation{comp}{Department of Astronomy, The Ohio State University, Columbus, OH, USA}
\mlforastroaffiliation{sch}{Center for Cosmology and AstroParticle Physics (CCAPP), The Ohio State University, Columbus, OH, USA}
\mlforastroaffiliation{tsinghua}{Department of Astronomy, Tsinghua University, Beijing, China}

\mlforastrocorrespondingauthor{Nesar Ramachandra}{nramachandra@anl.gov}

\mlforastrokeywords{Machine Learning, Cosmology, Galaxy, Spectra}

\vskip 0.3in
]



\printAffiliationsAndNotice{}  

\begin{abstract}
Pre-trained Large Language Models (LLMs) have revolutionized text processing, yet adapting Transformer-based neural networks to non-textual scientific modalities typically requires specialized architectures and extensive computational resources. We demonstrate that LLaMA-3.1-8B can be efficiently repurposed to predict galaxy redshifts from spectroscopic data through Low-Rank Adaptation (LoRA), achieving competitive performance while preserving its linguistic capabilities. Using only 16 GPU-hours and adapting 0.04\% of model parameters, our approach achieves a mean absolute error of 0.04 in redshift prediction while retaining over 85\% of performance on AstroBench and 89\% on general QA tasks from eval-harness. This minimal-effort adaptation—requiring only simple standard fine-tuning APIs—lowers barriers to entry for domain scientists and enables integrated agentic workflows where a single model handles both spectroscopic data for quantitative analysis and natural language for reasoning.
\end{abstract}

\section{Introduction}
\label{sec:intro}
Transformer-based models have revolutionized natural language processing through Large Language Models (LLMs) \citep{brown2020language, radford2019language, zhang2023llama}, leveraging the scalable transformer architecture's self-attention mechanism \citep{vaswani2017attention}. This mechanism's ability to capture long-range dependencies has enabled successful extensions beyond text to images \citep{dosovitskiy2020image}, graphs \citep{kipf2016semi}, and spectral data \citep{liu2021swin, fu2021transformers}.

In astronomy, transformers have shown promise for processing time series \citep[e.g.,][]{Pan2024a} and spectroscopic data \citep[e.g.,][]{Leung2024,Rozanski2025a}, where long-range correlations encode critical physical information. These models exhibit neural scaling laws, demonstrating potential for scaling to larger architectures \citep{Pan2024b,Rozanski2025b}. However, astronomical applications typically train specialized transformers from scratch, requiring computational resources and domain expertise. These models often employ custom tokenization schemes, specialized positional encodings, and domain-specific masking strategies—each requiring careful design and validation \citep{Rozanski2025a}.

These specialized models face several practical limitations. First, they cannot leverage the rapidly evolving LLM ecosystem, including optimized inference frameworks \cite{yuan2024llm}, quantization techniques \cite{zhao2024atom}, and deployment tools designed for text transformers. Second, astronomy-specific architectures often lack compatibility with fast inference systems like vLLM or TensorRT-LLM \cite{kwon2023efficient}, limiting their deployment at scale. Third, integrating these models into agentic workflows like \citep{moss2025ai} requires building custom interfaces between LLMs and domain-specific components, often increasing system complexity, maintenance burden, and high token consumption.  

This raises a fundamental question: can we repurpose existing pre-trained LLMs to process entirely new scientific modalities through efficient adaptation? Such approaches have started to gain attention in other fields, including chemistry \citep{jablonka2024}, material design \citep{Gruver2024Fine}, and protein design \citep{lv2024}, but have not been demonstrated in astronomy yet. Such an approach would lower the barrier to entry for astronomers while benefiting from mature LLM infrastructure. Crucially, any adaptation must preserve the model's original text processing and reasoning capabilities—the goal is augmentation, not replacement.

We demonstrate that LLaMA-3.1-8B, fine-tuned via Low-Rank Adaptation (LoRA) \citep{hu2021lora}, can effectively predict galaxy redshifts from spectroscopic data while retaining its language capabilities. This parameter-efficient approach shows that generic transformer models can serve as versatile scientific tools, processing both textual and spectroscopic modalities without requiring specialized architectures or extensive training from scratch.

\section{Dataset}



As a proof of concept, we chose galaxy redshift prediction—a fundamental cosmology task where accurate photometric redshifts enable large-scale structure studies \citep{newman2022photometric}. While focusing on this high-impact application, our approach should generalize to other spectroscopic tasks where sequential patterns encode physical information.
We compiled galaxy spectra from SDSS DR16, selecting galaxies (type = 3) with $0<z<0.50 < z < 0.5$ and dereddened $i<18$ to ensure nearby, luminous sources \citep{ahumada202016th}. The query retrieved identifiers, coordinates, redshifts, photometric measurements, and spectroscopic details. We obtained FITS format spectra from the SDSS Science Archive Server, handling data reduction differences between SDSS and eBOSS pipelines. After converting from logarithmic to linear wavelength scales and normalizing fluxes, we obtained 10,000 galaxy samples. With equal-frequency binning, we sample 3,000 galaxies for training that uniformly span the redshift range. A validation set of 1,000 galaxies across the full redshift range is reserved for unbiased evaluation.

\section{Methodology}

To adapt pre-trained LLMs for spectroscopic analysis, we must address two fundamental challenges: representing continuous spectral data in a format LLMs can process, and fine-tuning the model without sacrificing its linguistic capabilities.

\subsection{Tokenization}

A key challenge lies in tokenization. Specialized astronomy transformers \citep{Pan2024a,Rozanski2025a} train custom MLP-based tokenizers from scratch, learning optimal representations for spectral features. However, this approach requires modifying the model architecture and training pipeline—precisely the barriers we aim to avoid. 

Instead, we test whether standard LLM tokenizers can handle spectra with minimal adaptation. We serialize each flux value into digits using a configurable base representation with specified precision. For example, with base=10 and prec=2, the value $4.56\!\to\!\texttt{"4|5|6"}$ where the leading space indicates positive sign and ``|'' separates individual digits. Complete spectra become concatenated strings with ``\texttt{ ,}'' delimiting values. For instance, [4.56, 7.54, 11.2] becomes \texttt{"4|5|6 , 7|5|4 , 1|1|2|0 ,"}. The serialization handles signed values, removes leading zeros for variable-length representation, and includes proper separators for unambiguous parsing. Each input is prepended with \texttt{"Galaxy spectrum is rescaled and encoded to an input series:"} and target with \texttt{"Redshift: "}. With a total of 3,000 galaxies, this amounts to roughly 1.6M tokens. While suboptimal compared to learned tokenization, this approach requires zero architectural changes and tests the lower bound of what's achievable with minimal effort.

\subsection{Model Selection and Fine-tuning}

Having established a tokenization strategy, we selected a model balancing capability with accessibility. LLaMA-3.1-8B-Instruct represents an optimal trade-off: smaller models might lack the capacity to retain linguistic abilities while learning new modalities, while larger models exceed typical astronomy computing budgets. The 8B parameter scale provides sufficient capacity for multi-modal learning while remaining trainable on modest GPU clusters.

We employed Low-Rank Adaptation (LoRA), which decomposes weight updates into low-rank matrices $W + \Delta W = W + BA$ where $B \in \mathbb{R}^{d \times r}$ and $A \in \mathbb{R}^{r \times k}$ with $r \ll \min(d,k)$ \citep{hu2021lora}. This approach freezes the original weights $W$ while training only the compact matrices $A$ and $B$. LoRA has become the standard fine-tuning method across both open and proprietary models—OpenAI, Anthropic, and other providers offer LoRA-based fine-tuning APIs. While we use open-weight models for experimental flexibility, our approach translates directly to proprietary platforms where astronomers could upload their data and utilize built-in fine-tuning pipelines.

Using rank-8 adapters (3.4M parameters, $\sim$0.04\% of total model parameters), two training epochs require only 16 A100 GPU hours total—well within typical astronomy computing allocations. In addition, each galaxy training point occupies less than 7\% of the 8K context window of LLaMA-3. 

\begin{figure*}[ht]
    \centering
    \includegraphics[width=0.9\textwidth]{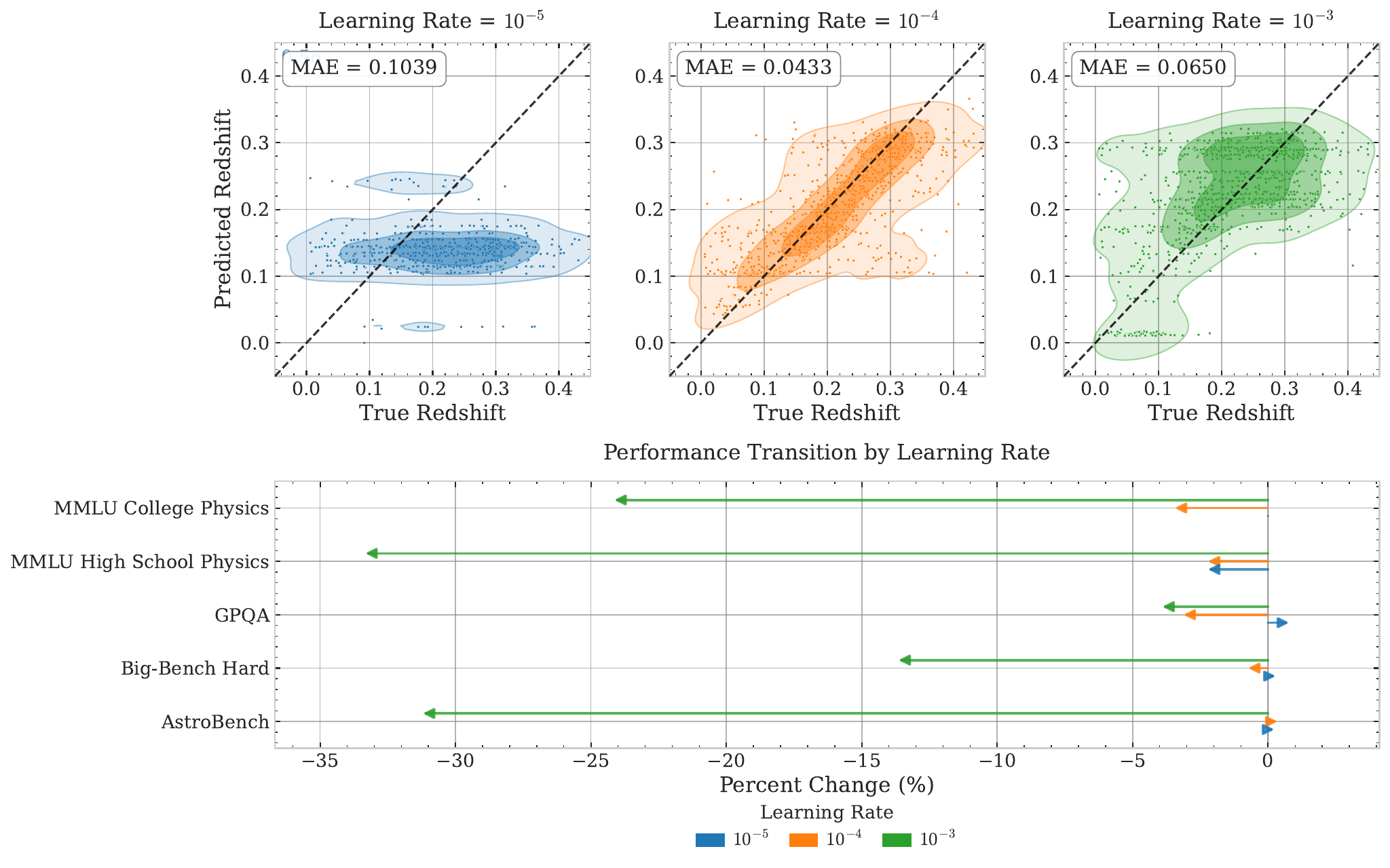}
    \vspace{-0.3cm}
    \caption{Trade-off between spectroscopic accuracy and language benchmark retention across learning rates. \textbf{Top:} Predicted vs.\ true redshifts for validation galaxies, with contours representing the full validation set of ~2,000 spectra and individual points shown for clarity. Learning rate $10^{-5}$ (left) preserves language capabilities but yields poor redshift predictions (MAE=0.104), while $10^{-4}$ (middle) achieves optimal spectroscopic accuracy (MAE=0.043) with acceptable language degradation, and $10^{-3}$ (right) shows intermediate spectroscopic performance (MAE=0.065) with substantial language degradation. \textbf{Bottom:} Percent change in language benchmarks after fine-tuning. The optimal learning rate $10^{-4}$ (orange) balances accurate redshift prediction with less than 15\% degradation in scientific reasoning tasks while maintaining strong general knowledge performance.}
    \label{fig:performance_tradeoff}
    \vspace{-0.3cm}
\end{figure*}

\section{Results}\label{results_sec}




After fine-tuning, our model serializes galaxy spectra flux values into tokens, generating responses as \texttt{"Redshift: [value]"}. We extract numerical predictions and compare against true spectroscopic redshifts for performance metrics. Beyond spectroscopic accuracy (mean absolute error, MAE), we evaluate pre-trained capability retention using eval-harness benchmarks \citep{eval-harness} and AstroBench, following \citet{Ting2025}.

The validation set comprised 20\% of galaxy spectra spanning the full redshift range. We analyze how learning rate, LoRA rank, training epochs, and dataset size affect the trade-off between modality adaptation and knowledge retention. Figure~\ref{fig:performance_tradeoff} illustrates the learning rate's critical role in balancing spectroscopic accuracy and language preservation. At $10^{-4}$, we achieve MAE=0.043 with less than 15\% decline in scientific reasoning and 89.4\% retention of general QA performance. The lowest rate ($10^{-5}$) preserves over 95\% of language capabilities but yields poor spectroscopic performance (MAE=0.104), while the highest rate ($10^{-3}$) improves redshift prediction (MAE=0.065) but degrades reasoning tasks by less than 20\%. Higher learning rates enable faster modality adaptation but disrupt the pre-trained representations more aggressively.

\begin{table*}[htbp]
\centering
\caption{Performance Comparison Across Different Fine-tuning Configurations}
\begin{tabular}{lcccccc}
\toprule
Learning Rate & LoRA Rank & Training size & Epochs & Redshift MAE$^\dagger$ & Scientific QA$^*$ & General QA$^*$ \\
\midrule
\multicolumn{7}{l}{\textbf{Varying Learning Rate} (rank = 8, epochs = 2, samples = 3,000)} \\
\midrule
$10^{-5}$ & 8 & 3,000 & 2 & 0.104 & \cellcolor{blue!20}96.5\% & \cellcolor{blue!20}95.1\% \\
\rowcolor{gray!20}
$\mathbf{10^{-4}}$ & 8 & 3,000 & 2 & \cellcolor{blue!20}\textbf{0.043} & \cellcolor{blue!12}\textbf{85.2\%} & \cellcolor{blue!12}\textbf{89.4\%} \\
$10^{-3}$ & 8 & 3,000 & 2 & \cellcolor{blue!12}0.065 & 76.2\% & 79.8\% \\
\midrule
\multicolumn{7}{l}{\textbf{Varying LoRA Rank} (rate = $10^{-4}$, epochs = 2, samples = 3,000)} \\
\midrule
$10^{-4}$ & 4 & 3,000 & 2 & 0.078 & \cellcolor{blue!20}87.8\% & \cellcolor{blue!20}91.2\% \\
\rowcolor{gray!20}
$10^{-4}$ & 8 & 3,000 & 2 & \cellcolor{blue!20}\textbf{0.043} & \cellcolor{blue!12}85.2\% & \cellcolor{blue!12}89.4\% \\
$10^{-4}$ & \textbf{16} & 3,000 & 2 & \cellcolor{blue!12}0.057 & 82.1\% & 86.7\% \\
\midrule
\multicolumn{7}{l}{\textbf{Varying Number of Epochs} (rate = $10^{-4}$, rank = 8, samples = 3,000)} \\
\midrule
$10^{-4}$ & 8 & 3,000 & 1 & 0.099 & \cellcolor{blue!20}87.9\% & \cellcolor{blue!20}91.5\% \\
\rowcolor{gray!20}
$10^{-4}$ & 8 & 3,000 & \textbf{2} & \cellcolor{blue!20}\textbf{0.043} & \cellcolor{blue!12}\textbf{85.2\%} & \cellcolor{blue!12}\textbf{89.4\%} \\
$10^{-4}$ & 8 & 3,000 & 3 & \cellcolor{blue!12}0.074 & 83.7\% & 88.1\% \\
\midrule
\multicolumn{7}{l}{$^\dagger$ Lower is better. $^*$ Higher is better. Fiducial configurations in \colorbox{gray!20}{gray}. Blue: \colorbox{blue!20}{best}, \colorbox{blue!12}{second}.} \\
\end{tabular}
\label{tab:performance_metrics}
\end{table*}

Beyond learning rate, our ablation studies reveal consistent patterns across other hyperparameters (Table~\ref{tab:performance_metrics}). Increasing LoRA rank from 4 to 16 improves redshift accuracy (MAE decreases from 0.078 to 0.057) by allowing more parameters to adapt, but higher ranks show diminishing returns while causing greater language degradation. This aligns with LoRA theory: higher ranks enhance expressive capacity but risk overwriting original knowledge. In addition, the training duration also matters: one epoch yields incomplete adaptation (MAE=0.099), whereas three epochs only marginally improve spectroscopy (MAE=0.074) at the cost of language degradation. These trade-offs highlight the importance of careful hyperparameter tuning to align scientific performance with language retention. 

Our chosen fiducial configuration—learning rate $10^{-4}$, LoRA rank 8, two epochs—achieves MAE = 0.043 while retaining over 85\% of original language capabilities, demonstrating an effective balance between spectroscopic accuracy and linguistic preservation. The 85.2\% retention of QA performance in AstroBench indicates that the model can still understand complex scientific concepts in astronomy and answer technical questions for integrated scientific workflows. This minimal degradation enables agentic analysis where the model processes raw spectra, derives physical parameters, and then reasons about the results in natural language. Below is a condensed conversation illustrating its dual capability:

\begin{questionbox}[boxcolor1]
\textbf{Step 1 - User provides spectrum:} [Galaxy spectrum data]\\
\textbf{Model prediction:} Photometric redshift: 0.315\\

\textbf{Step 2 - User:} What type of galaxy is expected at the redshift you just predicted? Which surveys could observe it?\\

\textbf{Model:} At redshift z=0.315, this is likely a massive elliptical galaxy. These are among the most massive and evolved galaxies, typically formed through mergers. The redshift corresponds to a lookback time of $\sim$3.8 billion years, during an active epoch of galaxy evolution.\\

Observable by: SDSS, DES, HST, ALMA (dust/gas), and future surveys like Euclid and LSST. The galaxy is receding at $\sim$84,000 km/s, with spectral features like H$\alpha$ shifted from 656nm to 863nm.
\end{questionbox}

For context, specialized spectroscopic redshift estimation methods can achieve lower MAE, even at sub-0.01 level \cite{bolton2012spectral}, suggesting our approach reaches competitive though not state-of-the-art performance. However, the value lies not in surpassing specialized methods but in demonstrating that general-purpose LLMs can achieve reasonable accuracy on scientific tasks while maintaining their broader capabilities, enabling truly integrated workflows where a single model handles the complete pipeline from raw data to scientific interpretation.

\section{Conclusion and Broader Impact}

This work demonstrates that pre-trained LLMs can be efficiently adapted to process non-textual scientific data through parameter-efficient fine-tuning. However, the tokenization, fine-tuning process, and the resulting performance may vary across the type of modalities. This approach contrasts sharply with training specialized models from scratch \citep[e.g.][]{Leung2024,Rozanski2025a}, which requires orders of magnitude more computational resources while sacrificing the ability to process natural language. Our findings have several important implications for scientific computing:

\textbf{Democratizing Access}: The approach substantially lowers barriers to entry for domain scientists. Rather than developing specialized architectures or training models from scratch, astronomers can leverage existing LLM infrastructure, fine-tuning APIs, and established deployment pipelines. 

\textbf{Enabling Integrated Workflows}: Our approach enables truly end-to-end scientific analysis within a single model. As demonstrated in our results, the same model that processes raw spectroscopic data to derive redshifts can immediately reason about the physical implications—discussing galaxy types, evolutionary stages, and observational strategies. This eliminates the need for complex interfaces between specialized components and enables more natural human-AI collaboration in scientific discovery.

\textbf{Revealing Transferable Representations}: Our results suggest that foundation models trained on text contain remarkably transferable representations applicable to diverse scientific modalities. The success of adapting a language model to spectroscopic analysis—achieving less than 15\% degradation in reasoning capabilities—implies that transformer pre-training captures general computational strategies for processing sequential information that transcend specific data types. While there are successful demonstrations of unified prediction frameworks like this in other specific scientific modalities \cite{hu20253dmolformer, zhang2025unigenx}, this area requires a deeper investigation. 

As LLM infrastructure continues its rapid advancement, parameter-efficient adaptation offers a practical path to democratize AI-driven scientific analysis. By building on existing foundation models rather than starting from scratch, the scientific community can benefit from ongoing improvements in language modeling while maintaining the flexibility to incorporate domain-specific data. This approach promises to accelerate scientific discovery by enabling models that can seamlessly navigate between quantitative analysis and conceptual reasoning—a capability increasingly essential for tackling complex agentic challenges.

\section*{Acknowledgements}
Work at Argonne National Laboratory is supported by UChicago Argonne LLC, Operator of Argonne National Laboratory. Argonne, a U.S. Department of Energy Office of Science Laboratory, is operated under Contract No. DE-AC02-06CH11357. The training is carried out on Swing, a GPU system at the Laboratory Computing Resource Center (LCRC) of Argonne National Laboratory. This material is based upon work supported by Laboratory Directed Research and Development (LDRD) funding
from Argonne National Laboratory, provided by the Director, Office of Science, of the U.S. Department of Energy under Contract No. DEAC02-06CH11357. YST is supported by the National Science Foundation under Grant No. AST-2406729. AZ and SH are supported by the U.S. Department of Energy, Office of Science, Office of High Energy Physics, High Energy Physics Center for Computational Excellence (HEP-CCE) at Argonne National Laboratory under B\&R KA2401045.






\bibliography{biblio}
\bibliographystyle{icml2025}

\newpage
\appendix
\onecolumn




\end{document}